\documentclass[preprint]{aastex}
\usepackage{booktabs}

\begin{document}

\title{Reconnection-driven Decaying Pulsations Modulated by Slow Magnetoacoustic Waves}

\author{Dong~Li$^{1,2}$, Jianping~Li$^{1}$, and Haisheng~Ji$^{1}$}
\affil{$^1$Purple Mountain Observatory, Chinese Academy of Sciences, Nanjing 210023, China \\
       $^2$Sate Key Laboratory of Solar Activity and Space Weather, National Space Science Center, Chinese Academy of Sciences, Beijing 100190, China
     }
     \altaffiltext{}{Correspondence should be sent to: lidong@pmo.ac.cn}
\begin{abstract}
Decaying pulsations have been simultaneously detected in the
low-energy X-rays of solar/stellar flares, which are supposed to be
associated with standing slow magnetoacoustic or kink-mode waves.
The physical mechanism behind rapidly decaying remains unknown. We
present the detection of quasi-periodic pulsations (QPPs) with
rapidly decaying in high-energy emissions produced in two major
flares on 10 January and 14 May 2024. Using empirical mode
decomposition, decaying QPPs are identified in hard X-ray and
microwave emissions during the flare impulsive phase, suggesting a
process of oscillatory magnetic reconnection. The quasi-periods and
decay times are determined by a damped harmonic function, which are
approximately 177$\pm$8~s (249$\pm$25~s) and 118$\pm$4~s
(124$\pm$5~s), respectively. The restructured X-ray images reveal
double footpoints connected by hot flare loops. Their phase speeds
are estimated to about 400~km~s$^{-1}$ and 670~km~s$^{-1}$, both
below the local sound speed in high-temperature plasmas, indicating
the presence of slow-mode waves in hot flare loops. We perform
coronal diagnostics based on standing slow-mode waves and derive key
physical parameters, including the polytropic index, the thermal
ratio, viscous ratio and radiation ratio, which are consistent with
previous results. Our observations support that the decaying QPPs
are triggered by oscillatory magnetic reconnection that is modulated
by standing slow magnetoacoustic waves, with their rapid decay
attributable to a co-effect of viscous damping and localized
magnetic reconnection rate.
\end{abstract}


\section{Introduction}
Quasi-periodic pulsation (QPP) is a very common phenomenon in the
process of solar/stellar eruption activity. It represents a
manifestation of the dynamic process in flare source region, which
can be used to directly diagnose the physical parameters of these
sources \citep[e.g.,][]{Tan08}. A typical QPP is characterized by a
group of successive and repetitive pulsations in the time-intensity
curves \citep[e.g.,][a reference therein]{Zimovets21}. The QPP
feature has been detected across the full electromagnetic spectrum
in solar flare emissions, i.e., radio, white light, ultraviolet
(UV), Ly$\alpha$, extreme ultraviolet (EUV), soft and hard X-rays
(SXR/HXR), and even $\gamma$-rays
\citep[e.g.,][]{Kolotkov18,Shen22,Collier24,Li24a,Li24b,Li25j,Kim25,Lim25,Shi25,Song25,OHare25,Zhang25}.
The quasi-periodic behaviors, which are usually regarded as
quasi-periodic oscillations (QPOs) in intensity curves, have been
extensively documented in radio, white light, and X-ray emissions
during stellar flares \citep{Hatt23,Pelisoli23}. The
topic of flare QPPs is important and of continued interest in the
astrophysics community, mainly due to their established connections
to particle acceleration, magnetohydrodynamic (MHD) waves, and
magnetic reconnection \citep[e.g.,][]{Inglis23}. The observed QPPs
are often explained by MHD waves in magnetic loops
\citep{Nakariakov20}, or they may be associated with nonthermal
particles periodically accelerated by a repetitive regime of
oscillatory magnetic reconnection \citep{Talbot24,Kumar25,Wang25}.
MHD waves can directly modulate the plasma temperature and
density in flare site \citep{Yuan15,Nakariakov19}, or they may
modulate the flare radiation by periodic variations of the angle
between the line of sight and the local magnetic field direction
\citep{Kohutova20}. Yet, MHD waves can quasi-periodically modulate the
efficiency and number of energetic electrons via the oscillatory
magnetic reconnection, leading to quasi-periodic accelerations and
precipitations of nonthermal electrons during solar flares
\citep{Jakimiec10,Chen15}.

Flare QPPs are detectable across all flare phases. The QPPs observed
in the pre-flare phase may serve as precursors of powerful flares,
and these seen in the impulsive phase may be used for diagnosing
periodic particle accelerations and oscillatory magnetic reconnection,
while those detected in the decay phase are usually associated with
MHD waves
\citep[e.g.,][]{Hayes16,Tan16,Chen19,Li21,Abramov24,Li25,Li25m}. A
well-known QPP feature observed during flare decay phases is termed
as SUMER oscillations. They are first reported by \cite{Wang02} from
the Solar Ultraviolet Measurements of Emitted Radiation (SUMER)
instrument. These oscillations are routinely detected in Doppler
shifts and SXR intensity curves of hot plasma loops
\citep{Wang03,Mariska06}. They are found to be excited by an energy
release at one footpoint of hot loops \citep{Kumar15}, and thus are
interpreted in terms of slow magnetoacoustic waves
\citep{Ofman02,Nakariakov04,Wang21}. SUMER oscillations frequently
exhibit the rapid-damping feature, and such characteristic is
consistent with decaying QPPs observed in SXR light curves during
solar and stellar flares \citep{Cho16}. The issue of rapidly damping
oscillations has attracted a remarkable attention since their
discovery, and multiple damping mechanisms have been proposed
\citep{Wang21}, i.e., non-adiabatic effects, wave-induced
imbalances, magnetic effects, wave-caused nonlinearity, and wave
leakages in the corona
\citep[e.g.,][]{Ofman02,Ofman22,Dem04,Selwa07,Provornikova18,Kolotkov19,Wang19,Prasad21}.

Decaying flare pulsations have been simultaneously reported in the
low-energy range of X-rays on the Sun and solar-like starts, which
are supposed to be related to the standing slow magnetoacoustic or
kink-mode waves in coronal loops \citep{Cho16}. In this article, we
present novel observations of rapidly decaying QPPs driven by
oscillatory magnetic reconnection in HXR and microwave emissions
during two solar flares. The multi-wavelength data provides
compelling evidence for modulation by standing magnetoacoustic waves
in hot flare loops. We also investigate the origin of rapidly
damping.

\section{Observations}
We conducted analysis of two major solar flares on 10 January and 14
May 2024, which are labeled as SF1 and SF2, respectively. They were
measured by various spaced-based instruments and ground-based
telescopes: the Macau Science Satellite-1 \citep[MSS-1;][]{Zhang23},
the Geostationary Operational Environmental Satellite (GOES), the
Large-Yield RAdiometer (LYRA), the Spectrometer/Telescope for
Imaging X-rays \citep[STIX;][]{Krucker20}, the Atmospheric Imaging
Assembly \citep[AIA;][]{Lemen12}, the Hard X-ray Imager
\citep[HXI;][]{Su19}, and the Expanded Owens Valley Solar Array
(EOVSA). The detailed parameters for the observational instruments
employed in this study are listed in table~\ref{tab1}.

\begin{table*}[ht]
\addtolength{\tabcolsep}{6pt} \centering \caption{Instruments and parameters employed in this study.} \label{tab1}
\begin{tabular}{cccccc}
\toprule
Instrument  &  Waveband     &  Cadence   & Description             &    Flare ID   \\
\midrule
MSS-1      &  1-24~keV      &   1~s      &  X-ray spectrum         &   SF1/SF2   \\
\midrule
           &  1-8~{\AA}     &   1~s      &  SXR flux               &   SF1/SF2   \\
GOES       &  3-25~keV      &   1~s      &  X-ray flux             &   SF1/SF2   \\
\midrule
LYRA       &  1-200~{\AA}   &   0.05~s   &  XUV flux               &   SF1/SF2   \\
\midrule
STIX       &  4-150~keV     &   1~s      &  X-ray spectrum/image   &   SF1/SF2   \\
\midrule
AIA        &  94/131{\AA}   &   24~s     &   EUV image             &   SF1/SF2   \\
\midrule
HXI        &  10-50~keV     &   1~s      &  X-ray spectrum/image   &     SF1    \\
\midrule
EOVSA      &  1-18~GHz      &   1~s      &  Radio spectrum         &     SF2    \\
\bottomrule
\end{tabular}
\end{table*}

\begin{figure}
\centering
\includegraphics[width=\linewidth,clip=]{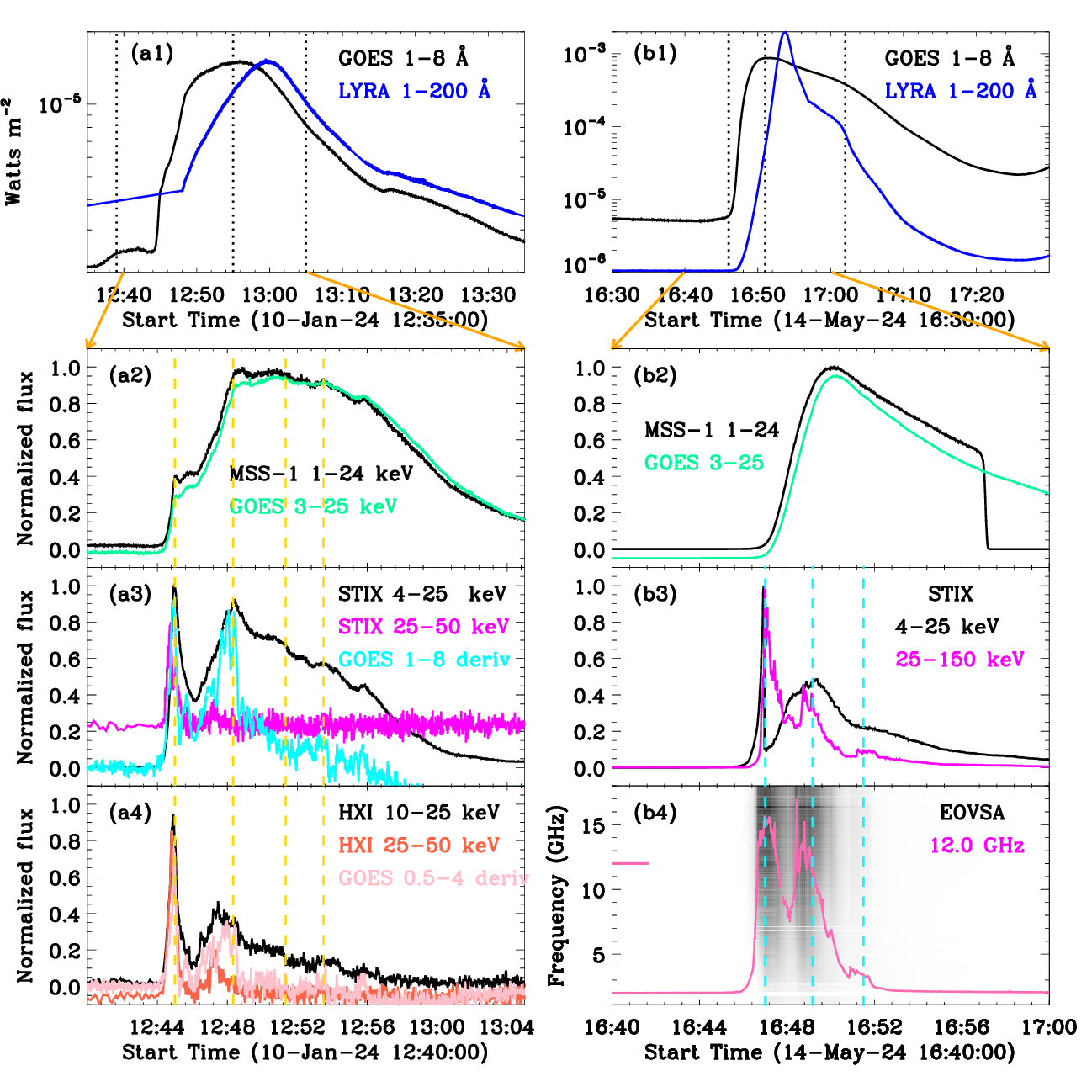}
\caption{Overview of two major flares on 10 January and 14 May 2024.
(a1 \& b1): Light curves recorded by GOES~1-8~{\AA} and
LYRA~1$-$200~{\AA}. The vertical dotted lines marks their start,
peak and stop times. (a2-a4): Normalized time series measured by
MSS-1, GOES, STIX, and HXI. (b2-b4): Normalized time series captured
by MSS-1, GOES, STIX, and EOVSA. The context image is the radio
dynamic spectrum measured by EOVSA. \label{over}}
\end{figure}

Figure~\ref{over}~(a1) \& (b1) show full-disk light curves in SXR
and X-ray ultraviolet (XUV) wavebands recorded by GOES~1-8~{\AA} and
LYRA~1-200~{\AA}, respectively. The SXR flux measurements suggest
that SF1 was an M1.4-class flare, which initiated at 12:39 UT,
peaked at 12:55 UT, and terminated at 13:05 UT; while SF2 was an
X8.7-class flare, which started at 16:46~UT, reached its maximum at
16:51~UT and stopped at 17:02~UT, as marked by the vertical dotted
lines. Panels~(a2)-(a4) present the SXR/HXR fluxes during the M1.4
flare measured by MSS-1, GOES, STIX and HXI. It should be
pointed out that the different light travel times to the Earth and
Solar Orbiter have been accounted for. That is, the STIX time is
modified to be consistent with the observations at the Earth. The
intensity curves at lower energy channels (i.e., $\leq$ 25~keV)
exhibit some small-scale pulsations superimposed on strong
background radiation. These pulsations are successive and rapidly
decaying, which may be regarded as rapidly decaying QPPs, as
indicated by the vertical dashed lines. The first peak
measured by MSS-1 and GOES is much weaker than that observed by STIX
and ASO-S/HXI, which maybe attributed to the different components.
That is, the first peak measured by MSS-1 and GOES is dominated by
the thermal radiation, and that observed by STIX and HXI is
dominated by nonthermal emissions. The GOES derived fluxes, which
can be regarded as the HXR radiation \citep[Neupert
Effect;][]{Neupert68,Li24}, reveal the similar morphology to the
STIX and HXI curves, as shown in Figure~\ref{over}~(a3) and (a4).
The energy windows as listed in GOES (a2) and STIX (a3) are very similar, but the corresponding light curves are so different, which may be attributed to different detectors of observed instruments. Panels~(b2)-(b4) show the SXR/HXR and microwave fluxes during the
X8.7 flare observed by MSS-1, GOES, STIX and EOVSA. The intensity
curves at higher energy channels (i.e., $>$ 25~keV) exhibit three
large-amplitude successive pulsations with rapidly decaying, as
indicated by the vertical dashed lines. Conversely, the intensity
curves at lower energy channels (i.e., $\leq$ 25~keV) lack analogous
feature, suggesting that the rapidly decaying QPPs are highly
associated with nonthermal electrons. The sharp decline in
panel~(b2) is mainly because that MSS-1 enters the shadow of Earth,
as it is a low-inclination satellite mission \citep{Zhang23}.

\section{Methods and Results}
\subsection{Periodicity analysis}
In order to distinguish decaying pulsations and background trend
from the original time series, the empirical mode decomposition
(EMD) technique \citep{Huang98} is applied. It can efficiently
decompose the observed intensity curve into a number of intrinsic
mode functions (IMFs), with the mode number typically approximating
log2($N$), where $N$ represents the number of data points of the
original signal. Unlike the other methods for detrending, EMD is
independent of any assumptions and has been successfully used for
detecting the QPPs signal during solar/stellar flares
\citep{Cho16,Kolotkov16}. The empirical background trend is
identified as those modes with periods exceeding a fraction of the
total signal duration, as denoted by a `cutoff' parameter, and it
returns a new set of modes that have periods shorter than the
cutoff. Those modes with rapidly damping and higher energy power
density are selected for subsequently analyzing.

\begin{figure}
\centering
\includegraphics[width=\linewidth,clip=]{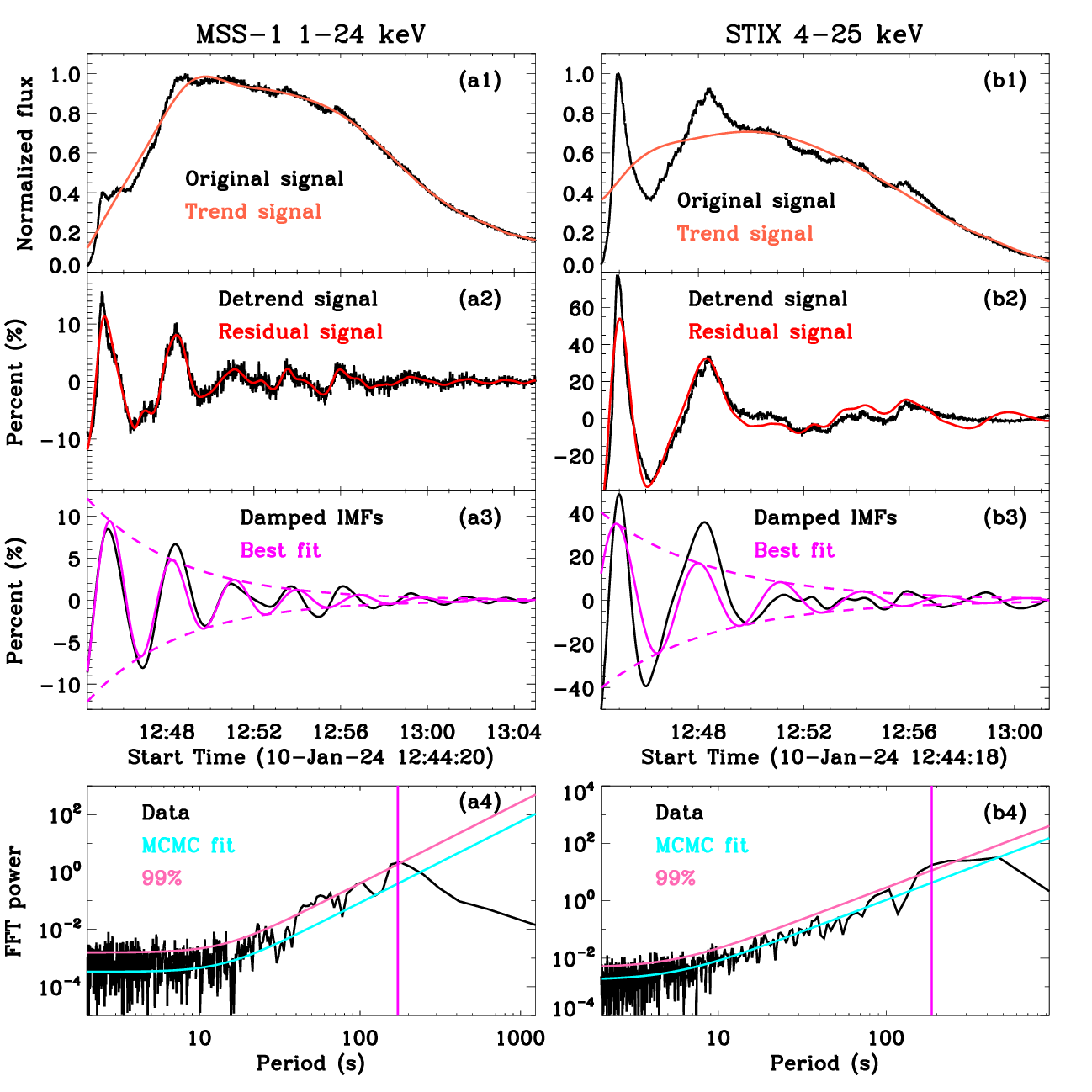}
\caption{EMD analysis of the M1.4 flare. (a1 \& a2): Normalized
intensity curves in wavebands of MSS-1~1-24~keV and STIX~4-25~keV.
The tomato line is the trend signal obtained from EMD . (b1 \& b2):
Detrended time series (original signal minus trend). The red curve
represents residual signal containing several IMFs. (c1 \& c2):
Damped oscillatory IMFs and their best fitting
with a damped harmonic function. (d1 \& d2): FFT power spectra of
detrended signals. The cyan line is MCMC-optimized fit, and the hot
pink line represents a confidence level at 99\%. The magenta
vertical line marks the period derived from least-squares fitting.
\label{emd1}}
\end{figure}

Figure~\ref{emd1} shows how we extracted decaying QPPs from
observational data using EMD. Panel~(a1) presents the normalized
time series in the waveband of MSS-1~1-24~keV, named as the original
signal. The trend signal, defined as a sum of IMFs at several lowest
frequencies, can effectively eliminate the steep decreasing pattern.
The detrended signal is derived from a subtraction of background
trends from the original signal. The residual signal, which
comprises several IMFs at middle frequencies, excludes both
background trends at lowest frequencies and micro-scale noises at
highest frequencies. This filtering is visually confirmed by the red
curve in panel~(a2), which exhibits smoothed oscillations devoid of
low- and high-frequency components. Panel~(a3) shows the decaying
QPP signal characterized by damped IMFs. Both detrended and damped
signals are normalized by the background trend, serving as the
modulation depth of decaying QPPs. The damped signal is subsequently
fitted by a damped harmonic function with a IDL code `mpfit.pro', as
specified in Eq.~\ref{eq1}. This IDL code utilizes a
Levenberg-Marquardt technique to solve the least-squares
minimization, and the best fit refers to a sum of the weighted
squared differences between the model and data is
minimized\footnote{http://cow.physics.wisc.edu/~craigm/idl/idl.html}.
The best-fitting curve for the decaying QPP is shown by the magenta
line, and the decaying trend is indicated by magenta dashed lines.
We noted that the fit looks pretty bad at later times/peaks,
because the later peaks do not belong to the decaying QPPs. On the
contrary, the fit is very well for the first four peaks, which are
regarded as the decaying QPPs. Table~\ref{tab2} lists key fitting
parameters, including fitting period ($P$), decay time ($\tau$), and
their ratio ($\tau/P$). Instrument-averaged quantities $\widehat{P}$
and $\widehat{\tau}$ utilize the standard deviation as an
uncertainty error.

\begin{equation}
\centering
 I (t) = I_0~sin(\frac{t-t_0}{P}+\phi)~e^{-\frac{t-t_0}{\tau}},
\label{eq1}
\end{equation}
\noindent Where $P$ and $\tau $ are the fitting period and decay
time, $I_0$, $t_0$, and $\phi$ denote to the initial amplitude,
start time, and phase of decaying QPPs.

\begin{table}[ht]
\addtolength{\tabcolsep}{5pt} \centering \caption{Fitted parameters for decaying QPPs.} \label{tab2}
\begin{tabular}{ccccccccccc}
\toprule
Flare ID  &   Instrument   &  $P$ (s)  & $\tau$ (s)  & $\tau/P$ & $\widehat{P}$ (s) &  $\widehat{\tau}$ (s)    \\
\midrule
     &    MSS-1 &  172   &  253     &  1.47     &                                      \\
SF1  &    STIX  &  187   &  258     &  1.38     &                                      \\
     &    GOES  &  180   &  271     &  1.51     &   177$\pm$8  &  249$\pm$25~s         \\
     &    HXI   &  170   &  213     &  1.25     &                                      \\
\midrule
     & EOVSA   &   121   &    127   &  1.05     &                                      \\
SF2 & STIX    &   115   &    120   &  1.04     &  118$\pm$4  &  124$\pm$5~s           \\
\bottomrule
\end{tabular}
\end{table}

It is well accepted that the IMFs decomposed by EMD always produce
quasi-periodic signals. Consequently, the quasi-periods derived from
damped IMFs require rigorous mathematical validation. The fast
Fourier transform (FFT) method is performed for detrended signals,
and the Bayesianbased Markov chain Monte Carlo (MCMC)
approach \citep{Guo23} is used to estimate the confidence interval
of a given value, and thereby, the Fourier peaks outside this
confidence interval are attributed to statistically significant
oscillatory processes of non-noise origin. Figure~\ref{emd1}~(a4)
presents the Fourier power spectrum in log-log space. The MCMC fit,
which consists of a power-law distribution ($P^{-\alpha}$) and a
constant ($C$), i.e., Eq.~\ref{eq2}, is applied to fit the power
spectral density, as shown by the cyan line. The hot pink line
represents the 99\% confidence interval, and the vertical line marks
the quasi-period derived from the least-squares fitting. The peak
period in the FFT spectrum is slightly deviated from the fitted
period, mainly due to the nonlinear nature of flare QPPs
\citep{Cho16}. Conversely, the FFT spectrum indicates a broad
distribution of quasi-periods, roughly aligning with the fitted
period. We also noted that several Fourier peaks exceeding
the confidence interval, indicating that multi-periods are
identified by the FFT method that is above the confidence interval.
In this case, only the decaying QPP is analyzed in detail.

\begin{equation}
\centering
 F (P) \propto P^{-\alpha} + C,
\label{eq2}
\end{equation}

Figure~\ref{emd1}~(b1)-(b4) show the EMD and FFT analysis results in
the energy range of STIX~4-25~keV. They reveal the same quasi-period
with rapidly decaying, confirming the existence of decaying QPP
during the M1.4 flare. The modulation depth of flare QPPs in
STIX~4-25~keV significantly exceeds that in MSS-1~1-24~keV, reaching
up to 50\%. Figure~\ref{emd_add} presents the EMD and FFT analysis
results in energy ranges of GOES~3-25~keV and HXI~10-25~keV. All
instruments show coherent decaying QPP patterns, and the modulation
depth in HXI~10-25~keV is much larger than that in GOES~3-25~keV.
The coordinated observations demonstrate the presence of decaying
QPPs pattern, because it can be simultaneously detected by different
instruments in different spacecrafts. Moreover, the decaying QPPs
pattern tends to appear in the higher X-ray range, since the higher
X-ray band ($\geq$10~keV) shows more pronounced oscillatory feature.

\begin{figure}
\centering
\includegraphics[width=\linewidth,clip=]{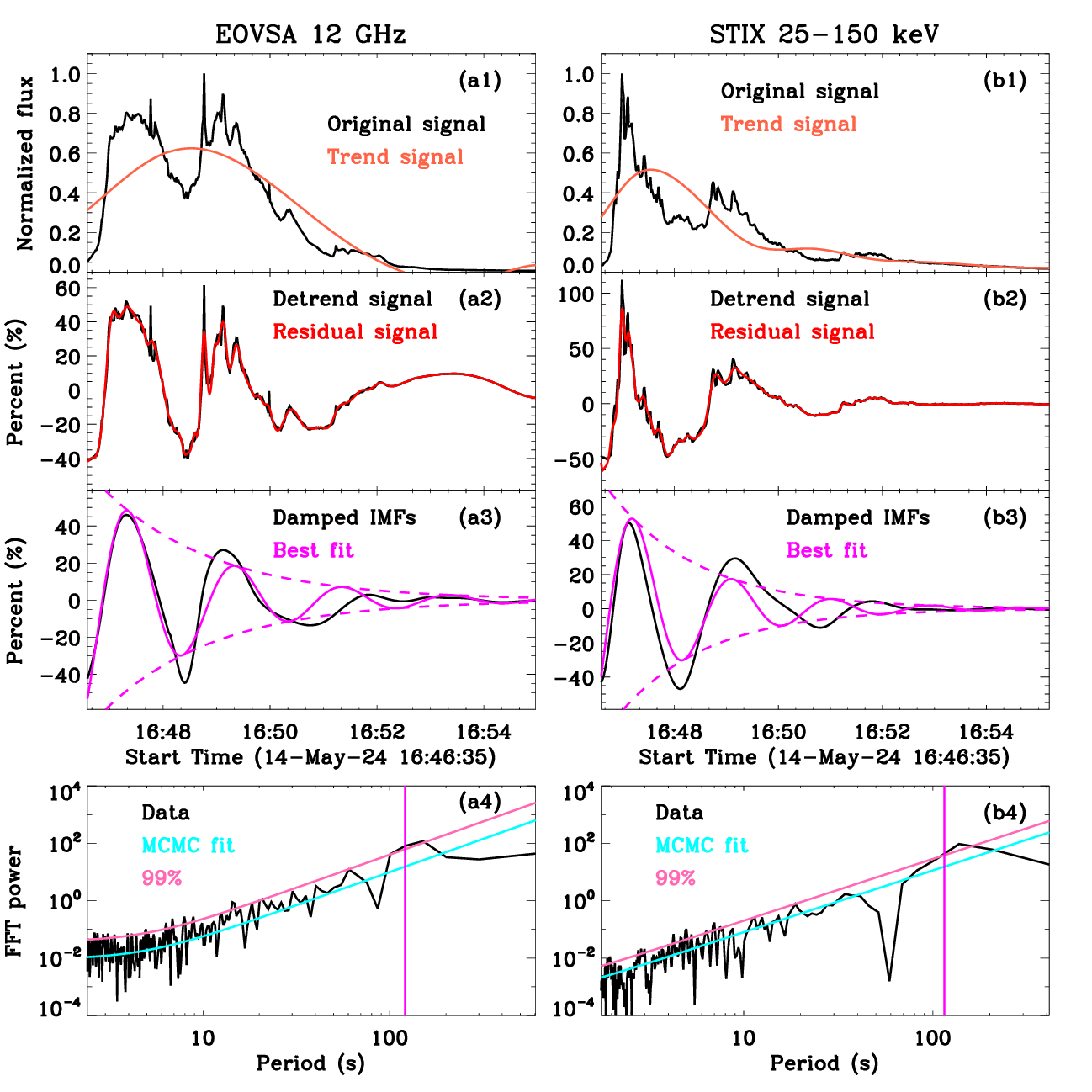}
\caption{Similar to Figure~\ref{emd1}, but the EMD and FFT analyses
are preformed for the X8.7 flare in channels of EOVSA~12~GHz and
STIX~25-150~keV. \label{emd2}}
\end{figure}

Based on the same EMD and FFT methods, we analyzed an X-class flare
in HXR and microwave emissions, as shown in Figure~\ref{emd2}.
Rapidly decaying QPPs are simultaneously detected in wavebands of
EOVSA~12~GHz and STIX~25-150~keV, with modulation depths
significantly exceeding those observed in the M-class flares, which
could be as high as 100\%. We noted that the fitted period from the
least-squares algorithm is substantially shorter than the FFT peak
period in the HXR radiation, primarily due to the significant
different durations of three pulsations. That is, the durations of
three successive pulsations become longer and longer, whereas the
fitting function inherently assumes a stationary period at the
beginning.

\subsection{Key parameter diagnosis}
\begin{figure}
\centering
\includegraphics[width=\linewidth,clip=]{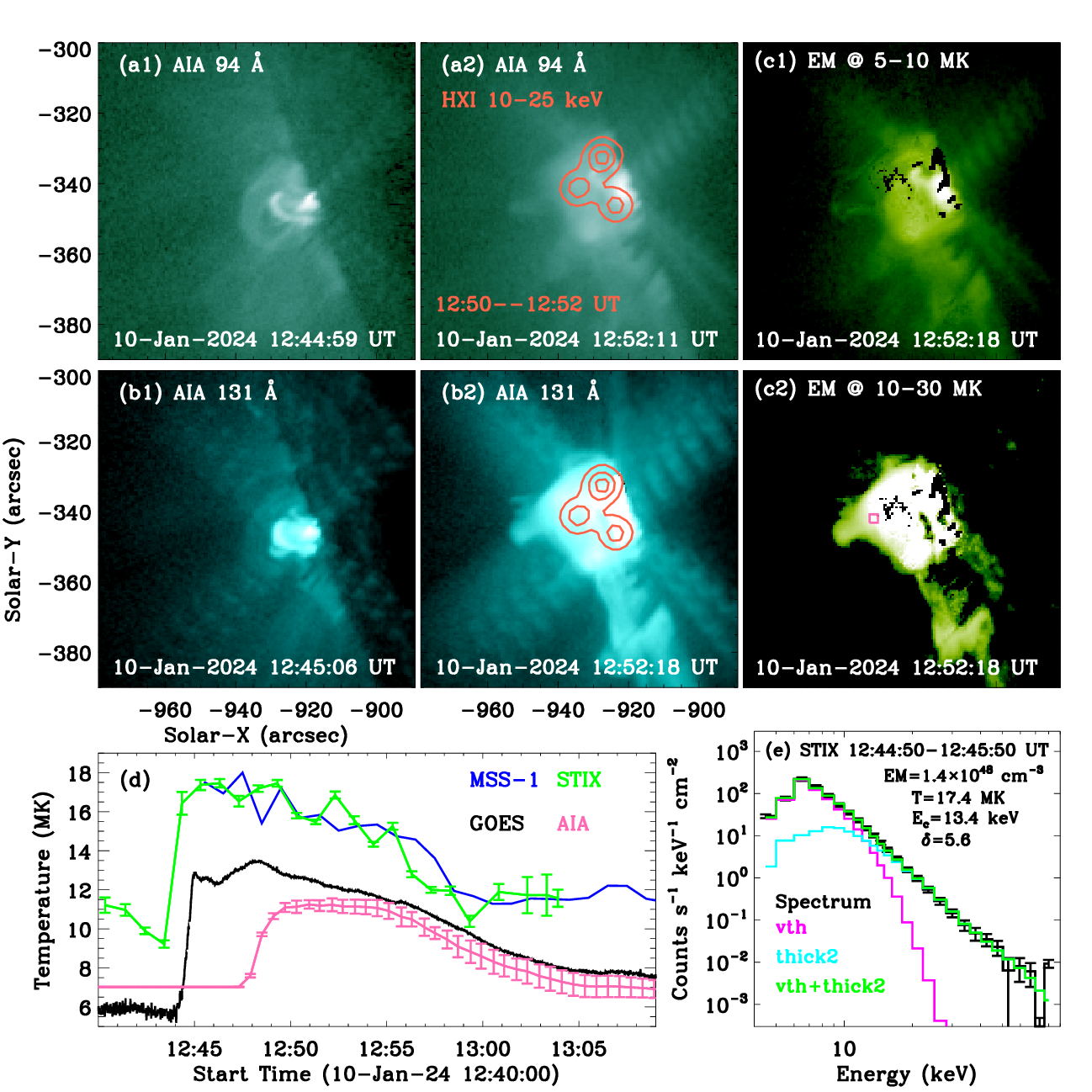}
\caption{(a1-b2): AIA EUV maps in wavelengths of 94~{\AA} and
131~{\AA} captured during the M1.4 flare, with a FOV of about
90\arcsec$\times$90\arcsec. The tomato contours represent the HXR
radiation measured by HXI at levels of 20\%, 50\%, and 80\%. (a3):
Narrow-band EM maps integrated over temperature ranges of 5-10~MK
and 10-30~MK. The hot pink box outlines the flare loop-top source,
used for estimating the plasma temperature. (d): Temporal evolutions
with error bars of the plasma temperature, as measured by MSS-1,
STIX, GOES, and AIA, respectively. (e): The X-ray spectrum with
error bars in the energy range of 4-80~keV, the fitted thermal and
nonthermal components, and their sum. Some fit parameters such as
EM, $T$, $E_{\rm c}$, and $\delta$ are labeled.  \label{img1}}
\end{figure}

Figure~\ref{img1}~(a1)-(b2) presents the spatial evolution of the
M1.4 flare during the decaying QPPs in wavebands of EUV and HXR.
High-temperature EUV maps measured by AIA~94~{\AA} and 131~{\AA}
reveal compact bright features during the initial phase of decaying
QPPs, which subsequently morph into loop-like structures as the
decaying pulsations progress. Double footpoints and loop top of the
hot loop can not be spatially resolved in EUV images due to the
saturation effect of AIA observations. Conversely, double footpoints
and loop-top sources are clearly identified in the HXR map, as
indicated by the tomato contours. Here, the HXR map is reconstructed
using the HXI\_CLEAN algorithm from the HXI observation with a pixel
scale of 2\arcsec. Under the assumption of a semi-circular flare
loop geometry \citep[cf.][]{Tian16}, the loop length ($L1$) derived
from the double footpoints measurement yields an approximate value
of 35.4~Mm.

The EUV maps in AIA~94~{\AA} and 131~{\AA} filter channels also
exhibit signatures of lower-temperature plasmas, manifested
primarily as continuum radiation contributions. Based on the AIA
observations in six EUV wavebands, the improved sparse-inversion
code developed by \cite{Su18} is used to compute the differential
emission measure (DEM) distribution at each pixel. Panels~(c1) and
(c2) show the narrow-band EM maps integrated over temperature ranges
of 5-10~MK and 10-30~MK, and the black patches are unsolvable points
due to saturation effects. The loop-like structures observed in the
high-temperature band further confirm the existence of hot flare
loops. The hot pink box outlines the loop-top source selected for
estimating the plasma temperature. This area is chosen based on its
distance from saturation-affected zones and its positioning at the
loop apex.

Figure~\ref{img1}~(d) depicts the temporal evolution of plasma
temperatures derived from multi-instrument observations. The blue
curve is derived from two SXR spectral features centered at about
3.9~keV (Ca) and 6.7~keV (Fe), obtained through comparison between
observational data (MSS-1) and theoretical atomic centroids
\citep[CHIANTI;][]{Li25p}. The green curve is calculated from the
STIX spectrum with the combination of a thermal component (vth) and
a thick-target component (thick2). To enhance the signal-to-noise
ratio and accumulate sufficient photons, both MSS-1 and STIX spectra
are integrated over 60~s. The data gap is attributed to the
insufficient photon counts. The black curve is determined by the
ratio of dual-channel GOES SXR fluxes under the assumption of an
isothermal plasma \citep{White05}. The hot pink curve represents a
EM-weighted temperature ($T_{\rm EM}$) at loop-top source,
calculated from AIA observations using Eq.~\ref{eq_dem}. Temperature
measurements derived from MSS-1 and STIX exhibit closely temporal
evolution, consistently exceeding values obtained from GOES and AIA
diagnostics. Because the temperature calculated from MSS-1
and STIX represents the extremely hot plasmas of the flare core,
while that measured from GOES is the isothermal temperature of
SXR-emitting plasmas in the entire Sun \citep{Li23}, and that
estimated from AIA is the EM-weighted temperature of EUV-emitting
plasmas in the flare loop. The different methods may measure the
various components in a multi-thermal plasma system. Therefore, the
highest temperature ($T_{\rm h}$) during the M1.4 flare is about
18~MK, and the average temperature ($\overline{T}$) at the loop top
is about 11~MK, which are determined from MSS-1 and AIA
observations, respectively. The number density ($n_{\rm e}$) at the
loop top is also calculated from DEM analysis with Eq.~\ref{eq_dem}.
Panel~(e) shows the fitted STIX spectrum for a representative time
interval, with some key parameters annotated including EM, plasma
temperature ($T$), cutoff energy ($E_{\rm c}$), and spectral index
($\delta$) of the injected electron flux. It can be seen than the
nonthermal spectrum is a bit soft, and the cutoff energy is as low
as 13.4~keV, suggesting that the decaying QPP may be dominated by
the HXR radiation mechanism.

\begin{equation}
\centering
 T_{\rm EM} = \frac{\int(DEM (T) \times T)~dT}{\int DEM~dT}, \quad
 n_{\rm e} = \sqrt{\frac{\int DEM (T)~dT}{w}}.
\label{eq_dem}
\end{equation}
\noindent Here DEM (T) denotes the temperature-dependent EM, $w$ is
the integration length along the line of sight and it is considered
to be equivalent to the full width at the half maximum of the hot
flare loop.

\begin{figure}
\centering
\includegraphics[width=\linewidth,clip=]{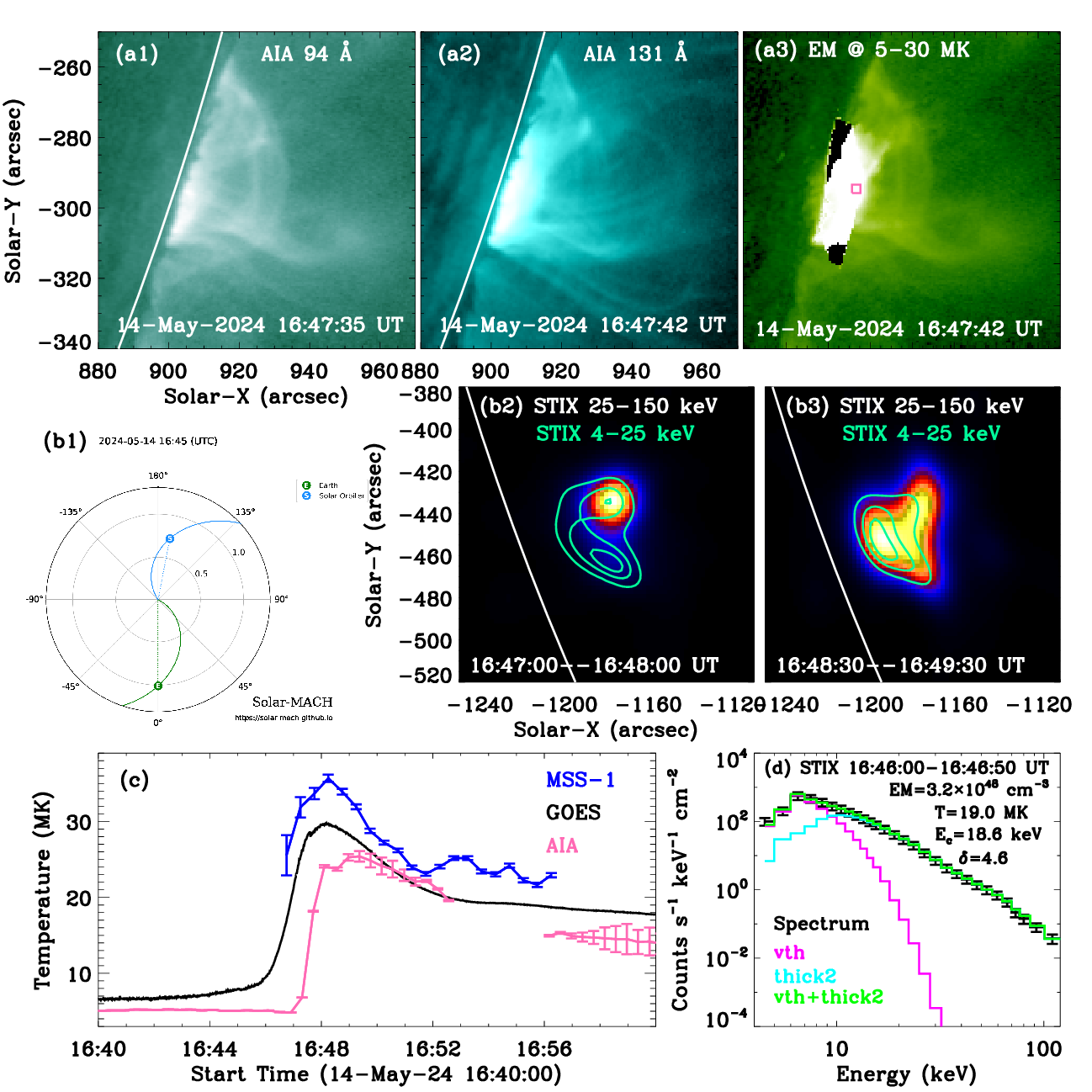}
\caption{(a1 \& a2): AIA EUV maps in wavelengths of 94~{\AA} and
131~{\AA} during the X8.7 flare, with a FOV of about
90\arcsec$\times$90\arcsec. (a3): Narrow-band EM map integrated over
the temperature range of 5-30~MK. The hot pink box outlines the
flare loop-top region, used to estimate the plasma temperature.
(b1): Sketch plot of the spatial locations of STIX and its
connection with the Sun and Earth. (b2-b3): Reconstructed STIX maps
in energy ranges of 25-150~keV, and 4-25 keV. (c): Temporal
evolutions with error bars of the plasma temperature, as measured by
MSS-1, GOES, and AIA. (d): The X-ray spectrum with error bars in the
energy range of 4-150~keV, the fitted thermal and nonthermal
components, and their sum. Some fit parameters such as EM, $T$,
$E_{\rm c}$, and $\delta$ are labeled. \label{img2}}
\end{figure}

Figure~\ref{img2} presents the same analysis for the X8.7 flare.
Panels~(a1)-(a3) show high-temperature EUV and narrow-band EM maps
obtained from AIA observations, revealing a strip-like feature that
may correspond to the hot flare loop. Panels~(b1)-(b3) present the
STIX locations and SXR/HXR maps during two flare pulsations. The
SXR/HXR maps are reconstructed from the STIX pixelated data using
the expectation-maximization (EM) algorithm
\citep{Massa19,Warmuth20}. The HXR radiation at 25-150~keV appears
at one footpoint and loop-top source, while the SXR/HXR radiation at
4-25~keV reveals a loop-like structure with double footpoints, as
indicated by spring green contours in panel~(b2). The flare loop
length ($L2$) is measured to about 39.6~Mm. Panel~(c) shows the
temporal evolution of plasma temperatures measured by MSS-1, GOES
and AIA. STIX undergoes detector attenuation during the powerful
flare, leaving only the initial X-ray spectrum usable for spectral
fitting. Pixel saturation occurred in AIA, resulting in data gaps
across specific EUV channels. MSS-1 and GOES show
temperature-evolution trends during the X8.7 flare, whereas MSS-1
consistently records higher temperatures than GOES, matching
observations from the M1.4 flare. Thus, the highest temperature
($T_{\rm h}$) during the X8.7 flare reaches about 36~MK, the average
temperature ($\overline{T}$) and the number density ($n_{\rm e}$) at
the loop top are about 23~MK and 4.7$\times$10$^{11}$~cm$^{-3}$,
respectively. Panel~(d) presents the spectrally fitted STIX data for
a selected time bin with key spectral parameters. The nonthermal
component of SF2 exhibits a harder spectrum and higher cutoff energy
than those of SF1.

\begin{table}[ht]
\addtolength{\tabcolsep}{3pt} \centering \caption{Physical parameters at the loop-top region during two flares.} \label{tab3}
\begin{tabular}{ccccccccccc}
\toprule
Flare ID &  $\overline{T}$ (MK)  &  $n_{\rm e}$ (cm$^{-3}$)  & $v_{\rm s}$ (km~s$^{-1}$)  & $\alpha$ &  $d$  &  $\epsilon$  & $r$     \\
\midrule
SF1  &      11     &  1.1$\times$10$^{11}$   &  500    & 1.07  &  0.0092   &  0.00041     &  0.096      \\
\midrule
SF2  &      23     &  4.7$\times$10$^{11}$   &  730    & 1.40  &  0.0098   &  0.00043     &  0.081       \\
\bottomrule
\end{tabular}
\end{table}

\section{Conclusion and Discussion}
We present a detection of decaying QPPs in the high-energy radiation
produced in two major solar flare on 10 January (M1.4) and 14 May
(X8.7) 2025. Based on STIX fluxes at the low-energy channel, the two solar flares could be regarded as double-peaked flares, which are commonly observed in fan-spine topology. However, both the STIX fluxes at the high-energy channel and microwave fluxes appear at least three main peaks, and they damped very quickly, which may be also regarded as decaying QPPs. The AIA images in Figure~4 and 5 show that both flares have a compact main flare site, but the longer loops are not seen in the nearby region. The observational feature might exclude the fact that the first two peaks are the driver. That is, the successive main peaks are regarded as flare QPPs. The decaying QPPs pattern is extracted from the
original time series using the EMD method \citep{Kolotkov16}, and
the quasi-period and decay time is determined by fitting a damped
harmonic function. The quasi-period is cross-validated through FFT
analysis combined with the MCMC approach \citep{Guo23}. The M1.4
flare exhibits an average period of 177$\pm$8~s in the lower X-ray
channels (i.e., $\leq$25~keV), with a decay time of 249$\pm$25~s.
Meanwhile, The X8.7 flare shows an average period of 118$\pm$4~s in
higher energy channels of HXR and microwave, with a corresponding
decay time of 124$\pm$5~s. These decaying QPPs exhibit large
modulation depths, i.e., $>$10\%, consistent with those measured in
flare QPPs from HXRs and $\gamma$-rays \citep{Nakariakov10,Li22}.
Such large modulations may indicate injections of energetic
electrons accelerated by magnetic reconnection during solar flares.

Flare QPPs are commonly attributed to MHD waves
\citep[e.g.,][]{Nakariakov20}. The decaying QPPs during the M1.4 and
X8.7 flares exhibit quasi-periods at about 177~s and 118~s. The
flare QPPs at such longer periods are difficult to be modulated by
sausage-mode waves, because they are fundamentally existence in the
quite thick and rather dense plasma loop. This limitation originates
from the propagation cutoff at lower wavenumbers, resulting in
characteristic periods of a few seconds in the solar corona
\citep{Nakariakov03,Tian16}. The phase speed ($v_{\rm ph}$) can be
determined from the loop length ($L$) and quasi-period ($P$), while
the adiabatic sound speed ($v_{\rm s}$) is estimated via the average
temperature ($\overline{T}$) of hot plasma loops, as shown in
Eq.~\ref{slow1}. The phase speeds in hot plasma loops of the M1.4
and X8.7 flares are estimated to approximately 400~km~s$^{-1}$ and
670~km~s$^{-1}$, respectively. The phase speeds are significantly
slower than the average speed of about 1300~km~s$^{-1}$ in a catalog
of kink oscillations \citep{Nakariakov20}, implying that the
decaying QPPs can not be triggered by kink-mode waves.

Based on the slow-mode MHD wave in Eq.~\ref{slow1}, the adiabatic
sound speeds in hot flare loops can be estimated to about
500~km~s$^{-1}$ and 730~km~s$^{-1}$, marginally higher than the
corresponding phase speeds. The observational fact suggests that the
flare QPPs may be associated with standing slow-mode MHD waves.
Subsequent coronal seismological analysis leverages these wave
properties to quantify plasma parameters. The polytropic index
($\alpha$) in both hot flare loops are estimated to about 1.07 and
1.40, consistent with previous measurements in the range of
1.05-1.58 \citep{Krishna18}, confirming their excitation by standing
slow magnetoacoustic waves. In the solar corona, flare QPPs
exhibiting large modulation depth (i.e., $>$20\%) can be directly
driven by slow magnetoacoustic waves, this occurs when the loop
sound crossing timescale ($\tau_{\rm s}$) exceeds the duration of
heating pulsations ($\triangle t_{\rm H}$) during magnetic
reconnection events \citep{Reale19}. In our case, the loop sound
crossing timescales of both major flares are precisely constrained
at about 132~s and 104~s during their respective temperature maxima.
Given the equivalence $\triangle t_{\rm H} \simeq P$, the flare QPPs
at quasi-periods of 177~s and 118~s can not be directly modulated by
slow magnetoacoustic waves, since they both show large modulation
depths. Flare QPPs predominantly appear during the impulsive phase
of major solar flares, which are clearly detected in HXR and
microwave emissions. Consequently, flare QPPs are attributed to
oscillatory magnetic reconnection modulated by standing slow MHD
waves -- specifically, the oscillation of magnetic reconnection is
directly triggered by slow-mode MHD waves, while flare QPPs are
indirectly modulated by standing slow magnetoacoustic waves.
In a word, the rate of magnetic reconnection is
quasi-periodically modulated by the slow-mode MHD wave, which
results into the quasi-periodic accelerations of the number of
precipitating energetic electrons, and thereby, to the QPPs in the
intensity curve of HXR and microwave emissions.

\begin{equation}
  v_{\rm s} = 152~\sqrt{\overline{T}({\rm MK})}, \quad
  v_{\rm ph} = \frac{2L}{P}, \quad
  \alpha \approx (\frac{v_{\rm ph}}{v_{\rm s}})^{2} \gamma, \quad
  \tau_{\rm s} \sim 5~\frac{L({\rm Mm})}{\sqrt{0.1~T_h({\rm MK})}} > \triangle t_{\rm H}.
  \label{slow1}
\end{equation}
\noindent Where, $v_{\rm s}$ denotes the adiabatic sound speed,
$\overline{T}({\rm MK})$ is the loop-averaged temperature in the
unit of MK, $\alpha$ and $\gamma$ represent the polytropic and
adiabatic indices, respectively.

Similarly to SUMER oscillations with rapidly damping
\citep{Ofman02,Wang02}, the flare QPPs exhibit rapid decay
characteristics and they require an effective decay mechanism.
Non-adiabatic effects, i.e., thermal conduction, compressive
viscosity, and optically-thin radiation, are considered in this
study. Because they are the most nominated mechanisms for damped
slow waves and have been studied intensively \citep[cf.][]{Wang21}.
Using Eq.~\ref{slow2}, the thermal ratio ($d$), viscous ratio
($\epsilon$), and radiation ratio ($r$) are quantitatively
constrained during two major flares, with detailed values presented
in table~\ref{tab3}. The thermal ratio for both solar flares is much
less than 0.1, implying a weak thermal conduction regime
\citep{Krishna14}. Concurrently, the damping contribution from
optically-thin radiation remains negligible, because it exclusively
operates at coronal temperatures below 5~MK \citep{Provornikova18}.
Conversely, the compressive viscosity has been demonstrated to
dominate over the damping in very-short coronal loops ($\sim$50~Mm)
at super-hot ($>$10~MK) temperature \citep{Prasad21,Ofman22}. This
model agrees with our observations, for instance, the loop-averaged
temperatures (lengths) are estimated to 11~MK (35.4~Mm) and 23~MK
(39.6~Mm), respectively. Therefore, the viscosity damping is
dominant over the effect of thermal conduction or optically-thin
radiative. It is difficult to evaluate the effect of wave-caused
misbalance between the cooling and heating processes proposed by
\cite{Kolotkov19}, since the misbalance timescales ($\tau_1$ and
$\tau_2$) are exclusively defined in the temperature range of
5-10~MK, which is lower than the loop-averaged temperature in our
case. Critically, the rapidly-decaying QPPs are detected in HXR and
microwave emissions, especially during the X8.7 flare. This
observational fact indicates that the evolution/oscillation of
magnetic fields plays an important role in rapidly decaying. That
is, the magnetic reconnection rate, which is highly related to
nonthermal electrons, determines the number of energetic electrons
\citep{Purkhart25,Wang25}. In summary, the rapidly-decaying
pulsations may result from the coupled effects of viscous damping
and magnetic reconnection rate modulation \citep{Dem04,Prasad21}.

\begin{equation}
  d = 4.93~(\frac{\overline{T}^{3/2}}{n_{\rm e}~P}), \quad
  \epsilon = 0.217~(\frac{\overline{T}^{3/2}}{n_{\rm e}~P}), \quad
  r = 2.7\times10^{-3}~(\frac{n_{\rm e}~P}{\overline{T}^{5/3}}).
  \label{slow2}
\end{equation}

Decaying QPPs have been observed in lower energy channels of SXR
fluxes (3-12~keV) during solar/stellar flares \citep{Cho16}, and
they statistically established a scaling law between the decay time
and quasi-period in solar QPPs, expressed as $\tau/P = 1.74\pm0.77$.
In our case, the average ratios between decay time and quasi-periods
are about 1.40 and 1.05 for the M1.4 and X8.7 flares, respectively.
Our results are systematically lower than the statistical scaling
law median of 1.74, yet remain within its established range of
0.99-2.51. This discrepancy may originate from our selection of
higher-energy X-ray bands: the decaying QPP in the M1.4 flare is
clearly observed in the X-ray range of 10-25~keV, and that in the
X8.7 flare persists in the HXR regime of 25-150~keV. Our
observations reveal an inverse correlation between the ratio of
$\tau/P$ and X-ray energy bands, that is, the ratio between decay
time and quasi-period decreases with higher observational energies
in X-rays. This preliminary finding requires validation through
large-sample statistical analysis, which will be the focus of our
studies in the future.

\acknowledgments The author would like to thank the referee
for his/her inspiring comments. This work is funded by NSFC under
grants 12250014, 12573057, the National Key R\&D Program of China
2022YFF0503002 (2022YFF0503000) and 2021YFA1600502 (2021YFA1600500),
the Strategic Priority Research Program of the Chinese Academy of
Sciences, Grant No. XDB0560000. The work is especially supported by
the Macao Foundation. D.~Li is also supported by the Specialized
Research Fund for State Key Laboratory of Solar Activity and Space
Weather. We thank the teams of MSS-1 HXI, STIX, GOES, AIA, LYRA,
EOVSA for their open data use policy.


\clearpage
\appendix
\renewcommand\thefigure{\Alph{section}\arabic{figure}}
\setcounter{figure}{0}
\section{Appendix}

\begin{figure}
\centering
\includegraphics[width=\linewidth,clip=]{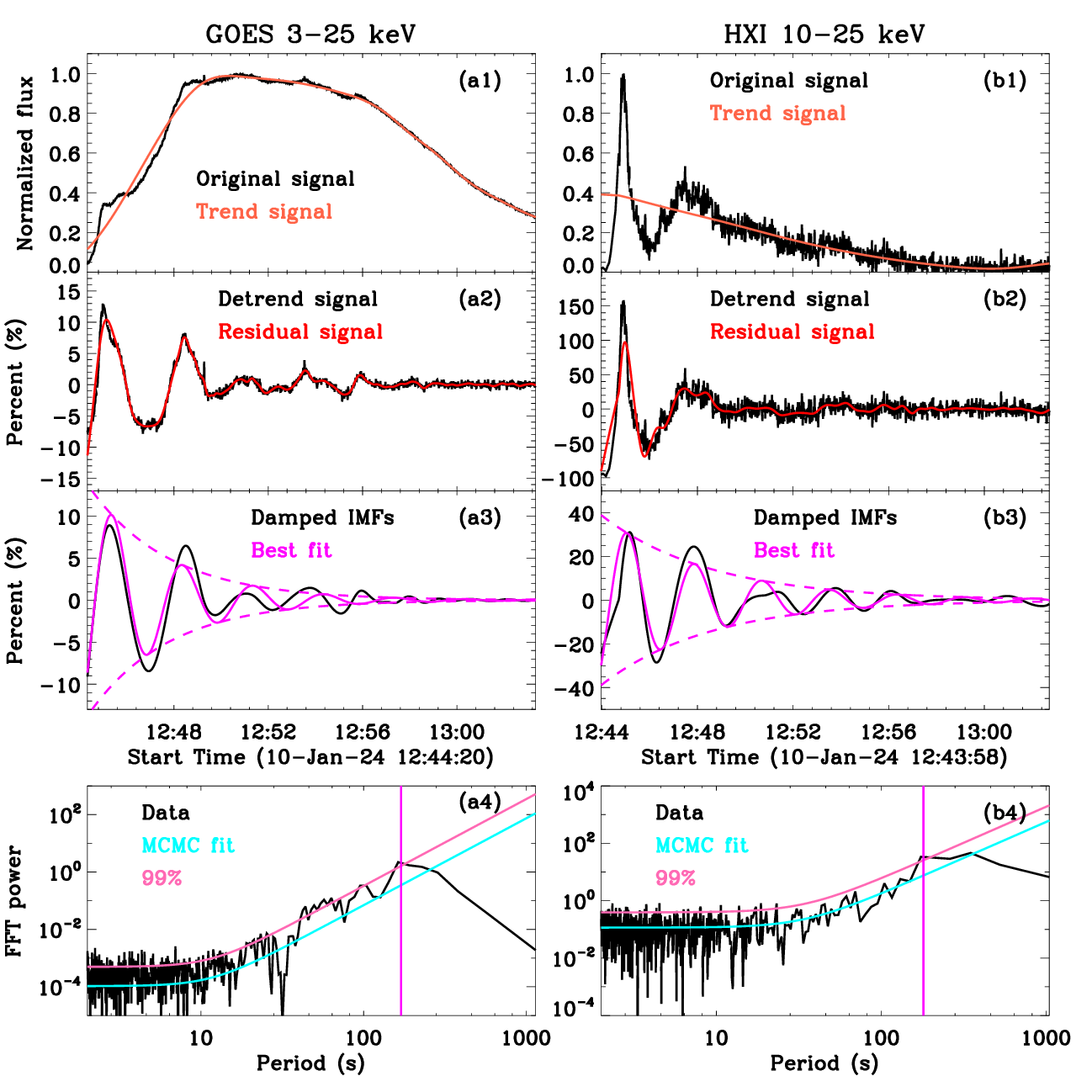}
\caption{Similar to Figure~\ref{emd1}, but the EMD and FFT analyses
are preformed for the M1.4 flare in wavebands of GOES~3-25~keV and
HXI~10-25~keV. (a1 \& a2): Normalized intensity curves, where the
tomato curve denotes the EMD-derived trend component. (b1 \& b2):
Detrended signal (original minus trend), with the red curve showing
residual IMFs. (c1 \& c2): Damped oscillatory IMFs and their
harmonic fits. (d1 \& d2): FFT power spectra of detrended signals.
The cyan line indicates MCMC-optimized fit, the hot pink line
represents a confidence level at 99\%, and the magenta vertical line
marks the least-squares fitted period. \label{emd_add}}
\end{figure}


\begin{thebibliography}{}
\bibitem[Abramov-Maximov \& Bakunina(2024)]{Abramov24} Abramov-Maximov, V.~E. \& Bakunina, I.~A.\ 2024, 64, 7, 1054. doi:10.1134/S0016793224700087
\bibitem[Collier et al.(2024)]{Collier24} Collier, H., Hayes, L.~A., Yu, S., et al.\ 2024, \aap, 684, A215. doi:10.1051/0004-6361/202348652
\bibitem[Chen et al.(2015)]{Chen15} Chen, B., Bastian, T.~S., Shen, C., et al.\ 2015, Science, 350, 6265, 1238. doi:10.1126/science.aac8467
\bibitem[Chen et al.(2019)]{Chen19} Chen, X., Yan, Y., Tan, B., et al.\ 2019, \apj, 878, 2, 78. doi:10.3847/1538-4357/ab1d64
\bibitem[Cho et al.(2016)]{Cho16} Cho, I.-H., Cho, K.-S., Nakariakov, V.~M., et al.\ 2016, \apj 830, 2, 110. doi:10.3847/0004-637X/830/2/110
\bibitem[De Moortel \& Hood(2004)]{Dem04} De Moortel, I. \& Hood, A.~W.\ 2004, \aap, 415, 705. doi:10.1051/0004-6361:20034233
\bibitem[Huang et al.(1998)]{Huang98} Huang, N.~E., Shen, Z., Long, S.~R., et al.\ 1998, Proceedings of the Royal Society of London Series A, 454, 1971, 903. doi:10.1098/rspa.1998.0193
\bibitem[Guo et al.(2023)]{Guo23} Guo, Y., Liang, B., Feng, S., et al.\ 2023, \apj, 944, 16. doi:10.3847/1538-4357/acb34f
\bibitem[Hatt et al.(2023)]{Hatt23} Hatt, E., Nielsen, M.~B., Chaplin, W.~J., et al.\ 2023, \aap, 669, A67. doi:10.1051/0004-6361/202244579
\bibitem[Hayes et al.(2016)]{Hayes16} Hayes, L.~A., Gallagher, P.~T., Dennis, B.~R., et al.\ 2016, \apjl, 827, 2, L30. doi:10.3847/2041-8205/827/2/L30
\bibitem[Inglis et al.(2023)]{Inglis23} Inglis, A., Hayes, L., Guidoni, S., et al.\ 2023, \baas. doi:10.3847/25c2cfeb.55d6b861
\bibitem[Jakimiec \& Tomczak(2010)]{Jakimiec10} Jakimiec, J. \& Tomczak, M.\ 2010, \solphys, 261, 2, 233. doi:10.1007/s11207-009-9489-4
\bibitem[Kim et al.(2025)]{Kim25} Kim, S., Lee, J., Park, S.-H., et al.\ 2025, \apj, 984, 1, 39. doi:10.3847/1538-4357/adc116
\bibitem[Kohutova et al.(2020)]{Kohutova20} Kohutova, P., Verwichte, E., \& Froment, C.\ 2020, \aap, 633, L6. doi:10.1051/0004-6361/201937144
\bibitem[Kolotkov et al.(2016)]{Kolotkov16} Kolotkov, D.~Y., Anfinogentov, S.~A., \& Nakariakov, V.~M.\ 2016, \aap, 592, A153. doi:10.1051/0004-6361/201628306
\bibitem[Kolotkov et al.(2018)]{Kolotkov18} Kolotkov, D.~Y., Pugh, C.~E., Broomhall, A.-M., et al.\ 2018, \apjl, 858, 1, L3. doi:10.3847/2041-8213/aabde9
\bibitem[Kolotkov et al.(2019)]{Kolotkov19} Kolotkov, D.~Y., Nakariakov, V.~M., \& Zavershinskii, D.~I.\ 2019, \aap, 628, A133. doi:10.1051/0004-6361/201936072
\bibitem[Krishna Prasad et al.(2014)]{Krishna14} Krishna Prasad, S., Banerjee, D., \& Van Doorsselaere, T.\ 2014, \apj, 789, 2, 118. doi:10.1088/0004-637X/789/2/118
\bibitem[Krishna Prasad et al.(2018)]{Krishna18} Krishna Prasad, S., Raes, J.~O., Van Doorsselaere, T., et al.\ 2018, \apj, 868, 2, 149. doi:10.3847/1538-4357/aae9f5
\bibitem[Kumar et al.(2015)]{Kumar15} Kumar, P., Nakariakov, V.~M., \& Cho, K.-S.\ 2015, \apj, 804, 1, 4. doi:10.1088/0004-637X/804/1/4
\bibitem[Kumar et al.(2025)]{Kumar25} Kumar, P., Karpen, J.~T., \& Dahlin, J.~T.\ 2025, \apj, 980, 158. doi:10.3847/1538-4357/ada293
\bibitem[Krucker et al.(2020)]{Krucker20} Krucker, S., Hurford, G.~J., Grimm, O., et al.\ 2020, \aap, 642, A15. doi:10.1051/0004-6361/201937362
\bibitem[Lemen et al.(2012)]{Lemen12} Lemen, J.~R., Title, A.~M., Akin, D.~J., et al.\ 2012, \solphys, 275, 17. doi:10.1007/s11207-011-9776-8
\bibitem[Li et al.(2021)]{Li21} Li, D., Ge, M., Dominique, M., et al.\ 2021, \apj, 921, 179. doi:10.3847/1538-4357/ac1c05
\bibitem[Li \& Chen(2022)]{Li22} Li, D. \& Chen, W.\ 2022, \apjl, 931, 2, L28. doi:10.3847/2041-8213/ac6fd2
\bibitem[Li et al.(2023)]{Li23} Li, D., Warmuth, A., Wang, J., et al.\ 2023, Research in Astronomy and Astrophysics, 23, 9, 095017. doi:10.1088/1674-4527/acd592
\bibitem[Li et al.(2024a)]{Li24a} Li, D., Li, J., Shen, J., et al.\ 2024a, \aap, 690, A39. doi:10.1051/0004-6361/202450622
\bibitem[Li et al.(2024b)]{Li24b} Li, D., Wang, J., \& Huang, Y.\ 2024b, \apjl, 972, L2. doi:10.3847/2041-8213/ad6cde
\bibitem[Li et al.(2024)]{Li24} Li, D., Dong, H., Chen, W., et al.\ 2024, \solphys, 299, 5, 57. doi:10.1007/s11207-024-02299-7
\bibitem[Li(2025a)]{Li25} Li, D.\ 2025a, \aap, 695, L4. doi:10.1051/0004-6361/202453613
\bibitem[Li(2025b)]{Li25m} Li, D.\ 2025b, \mnras, 542, 1, L48. doi:10.1093/mnrasl/slaf066
\bibitem[Li et al.(2025a)]{Li25j} Li, D., Yuan, D., Yan, J., et al.\ 2025a, Journal of Geophysical Research (Space Physics), 130, 4, e2025JA033772. doi:10.1029/2025JA033772
\bibitem[Li et al.(2025b)]{Li25p} Li, J., Yang, X., Li, D., et al.\ 2025b, Earth and Planetary Physics, 9, 3, 740. doi:10.26464/epp2025036
\bibitem[Lim et al.(2025)]{Lim25} Lim, D., Van Doorsselaere, T., Berghmans, D., et al.\ 2025, \aap, 698, A65. doi:10.1051/0004-6361/202554587
\bibitem[Mariska(2006)]{Mariska06} Mariska, J.~T.\ 2006, \apj, 639, 1, 484. doi:10.1086/499296
\bibitem[Massa et al.(2019)]{Massa19} Massa, P., Piana, M., Massone, A.~M., et al.\ 2019, \aap, 624, A130. doi:10.1051/0004-6361/201935323
\bibitem[Nakariakov et al.(2003)]{Nakariakov03} Nakariakov, V.~M., Melnikov, V.~F., \& Reznikova, V.~E.\ 2003, \aap, 412, L7. doi:10.1051/0004-6361:20031660
\bibitem[Nakariakov et al.(2004)]{Nakariakov04} Nakariakov, V.~M., Tsiklauri, D., Kelly, A., et al.\ 2004, \aap, 414, L25. doi:10.1051/0004-6361:20031738
\bibitem[Nakariakov et al.(2010)]{Nakariakov10} Nakariakov, V.~M., Foullon, C., Myagkova, I.~N., et al.\ 2010, \apjl, 708, 1, L47. doi:10.1088/2041-8205/708/1/L47
\bibitem[Nakariakov et al.(2019)]{Nakariakov19} Nakariakov, V.~M., Kosak, M.~K., Kolotkov, D.~Y., et al.\ 2019, \apjl, 874, 1, L1. doi:10.3847/2041-8213/ab0c9f
\bibitem[Nakariakov \& Kolotkov(2020)]{Nakariakov20} Nakariakov, V.~M. \& Kolotkov, D.~Y.\ 2020, \araa, 58, 441. doi:10.1146/annurev-astro-032320-042940
\bibitem[Neupert(1968)]{Neupert68} Neupert, W.~M.\ 1968, \apjl, 153, L59. doi:10.1086/180220
\bibitem[Ofman \& Wang(2002)]{Ofman02} Ofman, L. \& Wang, T.\ 2002, \apjl, 580, 1, L85. doi:10.1086/345548
\bibitem[Ofman \& Wang(2022)]{Ofman22} Ofman, L. \& Wang, T.\ 2022, \apj, 926, 1, 64. doi:10.3847/1538-4357/ac4090
\bibitem[O'Hare et al.(2025)]{OHare25} O'Hare, A.~N., Bekker, S., Hayes, L.~A., et al.\ 2025, Journal of Geophysical Research (Space Physics), 130, 4, e2024JA033493. doi:10.1029/2024JA033493
\bibitem[Pelisoli et al.(2023)]{Pelisoli23} Pelisoli, I., Marsh, T.~R., Buckley, D.~A.~H., et al.\ 2023, Nature Astronomy, 7, 931. doi:10.1038/s41550-023-01995-x
\bibitem[Prasad et al.(2021)]{Prasad21} Prasad, A., Srivastava, A.~K., \& Wang, T.~J.\ 2021, \solphys, 296, 1, 20. doi:10.1007/s11207-021-01764-x
\bibitem[Provornikova et al.(2018)]{Provornikova18} Provornikova, E., Ofman, L., \& Wang, T.\ 2018, Advances in Space Research, 61, 2, 645. doi:10.1016/j.asr.2017.07.042
\bibitem[Purkhart et al.(2025)]{Purkhart25} Purkhart, S., Collier, H., Hayes, L.~A., et al.\ 2025, \aap, 698, A318. doi:10.1051/0004-6361/202554475
\bibitem[Reale et al.(2019)]{Reale19} Reale, F., Testa, P., Petralia, A., et al.\ 2019, \apj, 884, 2, 131. doi:10.3847/1538-4357/ab4270
\bibitem[Selwa et al.(2007)]{Selwa07} Selwa, M., Ofman, L., \& Murawski, K.\ 2007, \apjl, 668, 1, L83. doi:10.1086/522602
\bibitem[Shen et al.(2022)]{Shen22} Shen, Y., Yao, S., Tang, Z., et al.\ 2022, \aap, 665, A51. doi:10.1051/0004-6361/202243924
\bibitem[Shi et al.(2025)]{Shi25} Shi, F., Warmuth, A., Li, D., et al.\ 2025, \apjl, 982, 1, L6. doi:10.3847/2041-8213/adb9e0
\bibitem[Song et al.(2025)]{Song25} Song, D.-C., Dominique, M., Zimovets, I., et al.\ 2025, \apjl, 983, 2, L41. doi:10.3847/2041-8213/adc4e9
\bibitem[Su et al.(2018)]{Su18} Su, Y., Veronig, A.~M., Hannah, I.~G., et al.\ 2018, \apjl, 856, 1, L17. doi:10.3847/2041-8213/aab436
\bibitem[Su et al.(2019)]{Su19} Su, Y., Liu, W., Li, Y.-P., et al.\ 2019, Research in Astronomy and Astrophysics, 19, 163. doi:10.1088/1674-4527/19/11/163
\bibitem[Tan(2008)]{Tan08} Tan, B.\ 2008, \solphys, 253, 1-2, 117. doi:10.1007/s11207-008-9235-3
\bibitem[Tan et al.(2016)]{Tan16} Tan, B., Yu, Z., Huang, J., et al.\ 2016, \apj, 833, 206. doi:10.3847/1538-4357/833/2/206
\bibitem[Talbot et al.(2024)]{Talbot24} Talbot, J., McLaughlin, J.~A., Botha, G.~J.~J., et al.\ 2024, \apj, 965, 2, 133. doi:10.3847/1538-4357/ad2a5d
\bibitem[Tian et al.(2016)]{Tian16} Tian, H., Young, P.~R., Reeves, K.~K., et al.\ 2016, \apjl, 823, L16. doi:10.3847/2041-8205/823/1/L16
\bibitem[Wang et al.(2002)]{Wang02} Wang, T., Solanki, S.~K., Curdt, W., et al.\ 2002, \apjl, 574, 1, L101. doi:10.1086/342189
\bibitem[Wang et al.(2003)]{Wang03} Wang, T., Solanki, S.~K., Curdt, W., et al.\ 2003, \aap, 406, 1105. doi:10.1051/0004-6361:20030858
\bibitem[Wang \& Ofman(2019)]{Wang19} Wang, T. \& Ofman, L.\ 2019, \apj, 886, 1, 2. doi:10.3847/1538-4357/ab478f
\bibitem[Wang et al.(2021)]{Wang21} Wang, T., Ofman, L., Yuan, D., et al.\ 2021, \ssr, 217, 2, 34. doi:10.1007/s11214-021-00811-0
\bibitem[Wang et al.(2025)]{Wang25} Wang, Y., Ni, L., Cheng, G., et al.\ 2025, \apj, 987, 2, 148. doi:10.3847/1538-4357/addd15
\bibitem[Warmuth et al.(2020)]{Warmuth20} Warmuth, A., {\"O}nel, H., Mann, G., et al.\ 2020, \solphys, 295, 7, 90. doi:10.1007/s11207-020-01660-w
\bibitem[White et al.(2005)]{White05} White, S.~M., Thomas, R.~J., \& Schwartz, R.~A.\ 2005, \solphys, 227, 2, 231. doi:10.1007/s11207-005-2445-z
\bibitem[Yuan et al.(2015)]{Yuan15} Yuan, D., Van Doorsselaere, T., Banerjee, D., et al.\ 2015, \apj, 807, 1, 98. doi:10.1088/0004-637X/807/1/98
\bibitem[Zhang(2023)]{Zhang23} Zhang, K.\ 2023, Earth and Planetary Physics, 7, 1, 4. doi:10.26464/epp2023019
\bibitem[Zhang et al.(2025)]{Zhang25} Zhang, Y., Li, T., Teng, W., et al.\ 2025, \apjl, 985, 1, L1. doi:10.3847/2041-8213/add33b
\bibitem[Zimovets et al.(2021)]{Zimovets21} Zimovets, I.~V., McLaughlin, J.~A., Srivastava, A.~K., et al.\ 2021, \ssr, 217, 66. doi:10.1007/s11214-021-00840-9
\end{thebibliography}
\end{document}